# Magnetoresistance in Metallic Ferroelectrics

Jiesu Wang,[†,‖] Hongbao Yao,[†,‡,‖] Kuijuan Jin,*[,†,‡,§] Er-Jia Guo,[†,‡] Qinghua Zhang,[†] Chao Ma,[†,⊥] Lin Gu,[†,‡] Pazhanivelu Venkatachalam,[†] Jiali Zhao,[‡,Δ] Jiaou Wang,[Δ] Hassen Riahi,[†] Haizhong Guo[ḉ], Chen Ge,[†] Can Wang,[†,‡,§] and Guozhen Yang[†,‡]

**Affiliations**

[†] Beijing National Laboratory for Condensed Matter Physics, Institute of Physics, Chinese Academy of Sciences, Beijing 100190, China

[‡] University of Chinese Academy of Sciences, Beijing 100049, China

[§] Songshan Lake Materials Laboratory, Dongguan, Guangdong 523808, China

[⊥] College of Information Science and Engineering, Shandong Agricultural University, Tai'an 271000, China

[Δ] Beijing Synchrotron Radiation Facility, Institute of High Energy Physics, Chinese Academy of Sciences, Beijing 100039, China

[ḉ] School of Physical Engineering, Zhengzhou University, Zhengzhou, Henan 450001, China

**ABSTRACT**

Polar metals with ferroelectric-like displacements in metals have been achieved recently, half century later than Anderson and Blount's prediction. However, the genuine ferroelectricity with electrical dipolar switching has not yet been attained experimentally in the conducting materials, especially the ones possessing magnetic responses. Here we report the coexistence of ferroelectricity and magnetoresistance (MR) in the metallic PbNb$_{0.12}$Ti$_{0.88}$O$_3$ (PNTO) thin films. We found that the conducting and magnetic responses of PNTO films are highly asymmetric. Negative MR up to 50% is observed under an in-plane magnetic field; the MR switches to positive with the magnetic field applied parallel to the surface normal. Such unique behavior is attributed to the moving electron caused effective magnetic field which couples with the spins of electrons, which form a *dynamic multiferroic* state in the metallic PNTO. These findings break a path to multiferroic metal and offer a great potential to the multi-functional devices.

## 1. INTRODUCTION

Ferroelectric-like behavior in a metal, proposed by Anderson and Blount in 1965,[1] used to be believed irreconcilable because the mobile charges would neutralize the intrinsic polarization completely and screen the electric field,[2-5] leading to the failure of electrical polarization switching.[6,7] That is the reason why the discoveries of the ferroelectric-like structural transition at 140 K in the metallic $LiOsO_3$ and the polar metal $NdNiO_3$ films have generated a flurry of interest.[4,7,8] These materials maintain the metallic behavior and exhibit the long-range ordered dipoles owing to cooperative atomic displacements aligned from dipolar-dipolar interactions, similar to the conventional ferroelectric insulators.[8] Numerous efforts had been dedicated to achieving the polar metals by either creating the artificial asymmetric potential in the conductive systems or introducing free electrons into the conventional ferroelectrics,[2,9,10,11] including several structures we proposed.[5,12-14] However, to date, a clear ferroelectricity—that is, the clear existence of a switchable intrinsic electrical polarization—has not yet been attained in a metal experimentally, let alone any ferromagnetic property found in any polar metal. Although the magnetism in ferroelectric-metal has been proposed,[8] the magnetic property in the aforesaid systems is seldom researched.

Ferroelectric $PbTiO_3$ (PTO) possesses tetragonal crystal structure, strong ferroelectric polarization, and intrinsic diamagnetic,[15] with lattice parameters $a = b = 3.90$ Å, $c = 4.15$ Å at room temperature. It undergoes a ferroelectric phase transition from tetragonal (*4mm*) to cubic (*m3m*) symmetry at 763 K ($T_C$).[16] Its ferroelectricity originates from the displacement of Ti and O shifted, with respect to Pb, in the same direction but with different values.[17] Ren *et al* have regulated the magnetism of PTO from diamagnetism to ferromagnetism by doping ferromagnetic metal (Fe) into PTO nanocrystals, and explained by the interaction between ferric ions and oxygen vacancies.[15] In our former experimental research, we realized the coexistence of conducting and polarization in Nb-doped PTO films, and found that the $PbNb_xTi_{1-x}O_3$ films are insulating with x = 0.04, 0.06, and 0.08, until 0.12.[5] The magnetism in PNTO films has never been experimentally found. Here we report the newly discovered polar metallic PbNb$_{0.12}$Ti$_{0.88}$O$_3$ thin films exhibiting an extraordinary magnetoresistance under applied magnetic fields. The distinguishing ferroelectric feature with switchable electric polarization was proved by macroscopic ferroelectric hysteresis loops and piezoresponse force microscopy (PFM) measurement at *room* temperature. Large central atomic displacement and *P4mm* symmetry were confirmed by scanning transmission electron microscopy (STEM) and second harmonic generation (SHG) measurements, respectively. Asymmetric electrical conductivities and magnetic responses were observed by changing the direction of applied magnetic field with respect to the electrical polarization. We attribute the origin of the magnetoresistance to the effective magnetic field induced by the electron moving in a system with intrinsic ferroelectric polarization. Further contrast experiments demonstrated this mechanism. These results should stimulate both fundamental understanding and technical interests for the potential use of multiferroic devices.

## 2. EXPERIMENTAL SECTION

**2.1. Synthesis of PNTO films.** The PbNb$_{0.12}$Ti$_{0.88}$O$_3$ (PNTO) films were epitaxially grown on the (001)-oriented SrTiO$_3$ (STO) and Nb-doped SrTiO$_3$ (0.1 *wt*%, SNTO) substrates by a laser molecular-beam epitaxy system (Laser-MBE). All substrates were treated to form a step-and-terrace feature on the surface. During the film deposition, the substrate temperature was kept at 520 °C, the oxygen partial pressure is maintained at 8 Pa, and the laser energy density was fixed at ~1.2 J/cm$^2$. Before cooled down to room temperature, all samples were annealed *in situ* at the deposition condition for 20 minutes to ensure the stoichiometry. The thickness of PNTO film is ~100 nm, controlled by the deposition time. The STO and SNTO substrates were used for the electrical measurements in transverse and longitudinal geometry, respectively.

**2.2. X-ray Diffraction and Absorption Spectral Measurements.** X-ray diffraction (XRD) and X-ray diffractometry reciprocal space mapping (RSM) measurements were performed to identify the phase structure of PNTO films, using Rigaku SmartLab (9 kW) High-resolution (Ge 220 × 2) X-Ray Diffractometer with 1.5406 Å X-rays. X-ray absorption (XAS) measurements were performed at the 4B9B beamline in Beijing Synchrotron Radiation Facility with the ultrahigh vacuum chamber background pressure being $2 \times 10^{-10}$ Torr. All measurements were carried out at room temperature.

**2.3. Magnetic and Electrical Measurements.** The magnetoresistance and electrical characteristics of PNTO films were conducted using a Keithley 2400 source meter, combined with a Quantum Design Physical Properties Measurement System (PPMS) with the temperature ranging from room temperature to 10 K and the applied magnetic field sweeping in ±1 T.

**2.4. Scanning Transmission Electron Microscopy (STEM) Measurement.** TEM specimens were obtained by mechanically polishing to about 20 μm. Central parts of the specimens were further reduced by precision argon-ion milling, until transparent for electron beam. The atomic structure of the PNTO was characterized using an ARM-200CF (JEOL, Tokyo, Japan) transmission electron microscope operated at 200 keV and equipped with double spherical aberration (Cs) correctors. The collection angle of high-angle annular dark-field (HAADF) image is 90-370 mrad. All HAADF images were filtered using the HREM-Filters Pro/Lite released by HREM Research Inc. Atomic positions were determined by fitting with Moment Method & Contour using the CalAtom Software developed by Prof. Fang Lin.

**2.5. Second-Harmonic Generation (SHG) measurement and analysis.** Far-field SHG polarimetry measurement was performed in a typical reflection geometry shown as the inset of Figure 1d. The incident laser beam was generated by a Spectra Physics Maitai SP Ti:Sapphire oscillator with the central wavelength at 800 nm (~120 fs, 82 MHz). The incident laser power was attenuated to 70 mW before being focused on the films. Both incidence angle and reflection angle were fixed at 45°. The polarization direction $\varphi$ of the incident light field was rotated by a $\lambda/2$ wave plate driven by rotational motor. Generated second harmonic light field from the PNTO films as well as the surface and interface were firstly decomposed into *p*-polarized (*p*-out) and *s*-polarized (*s*-out) components by a polarizing beam-splitter. After spectrally filtered, the second-harmonic (SH) signals were detected by a photo-multiplier tube. The read-out sum of frequency-doubled photons $S_{2\omega}$ is proportional to the SHG response.[18-20] The polar plots were acquired through rotating the incident light polarization by $\varphi$ for *p* and *s* components of the SH signal.

As the simplest order of nonlinear optic process, SHG describes the model of frequency-doubled light wave emitted by the polarization $\vec{P}_{2\omega}$ induced by fundamental light wave, and the intensity of SHG response can usually be written as $I_{2\omega} \propto |\vec{P}_{2\omega}|^2$.[19] Under the coordinate shown as the inset of Figure 1d, the SHG susceptibility tensor for the *4mm* point group symmetry has a form of[20]

$$\overset{\leftrightarrow}{\chi}^2_{4mm} = \begin{pmatrix} 0 & 0 & 0 & 0 & \chi_{15} & 0 \\ 0 & 0 & 0 & \chi_{15} & 0 & 0 \\ \chi_{31} & \chi_{31} & \chi_{33} & 0 & 0 & 0 \end{pmatrix}. \quad (1)$$

The solid lines in Figure 1d are theoretical fittings of the experimental data using the expressions:

$$\begin{cases} S_{4mm,p-out} \propto \left\{ \left[ \frac{\sqrt{2}}{2} L^\omega_{yy} L^\omega_{zz} L^{2\omega}_{yy} \chi_{15} - \frac{1}{2} L^{2\omega}_{zz} \left( L^\omega_{yy} \right)^2 \chi_{31} - \frac{1}{2} L^{2\omega}_{zz} \left( L^\omega_{zz} \right)^2 \chi_{33} \right] \cos^2\varphi - B\sin^2\varphi \right\}^2 \\ S_{4mm,s-out} \propto \left[ \frac{\sqrt{2}}{2} L^{2\omega}_{zz} \left( L^\omega_{xx} \right)^2 \chi_{31} \sin 2\varphi \right]^2 \end{cases}, \quad (2)$$

which are derived by substituting the expression (1) to the sum of SHG photons. $L_{xx}$, $L_{yy}$, and $L_{zz}$ are the diagonal elements of the transmission Fresnel factor $\vec{L}$ in materials for fundamental and second-harmonic light. Their expressions can be found in our former research as Ref. [19].

**2.6. Piezoresponse Force Microscopy (PFM) and Ferroelectric Measurements.** The microscopic ferroelectric domains were written and read using a commercial atomic force microscope (AFM, Asylum Research MFP-3D). The Ti/Ir-coated Si cantilever (Olympus Electrilever) was used to collect and record the PFM images, which has a nominal spring constant of ~2 N/m and a tested free air resonance frequency of ~73 kHz.

Ferroelectric tester (Radiant Technologies, Premier II) was employed to acquire the ferroelectric hysteresis loops (*P-E* loops) of Au/PNTO/SNTO at ambient temperature under the capacitance configuration. In this capacitance component, the *P-E* loops were derived by subtracting one hysteresis loop from another inverse one to minimize the removable charges contribution on the polarization signals. The measured frequency was set to 1 kHz.

**2.7. Density Functional Theory (DFT) calculation**. DFT calculations were performed within the generalized gradient approximation (GGA) with the PBE exchange and correlation functional as implemented in the Vienna *ab initio* simulation package (VASP).[21,22] The projector augmented wave method (PAW) was used with the following electronic configurations: $5d^{10}6s^26p^2$ (Pb), $3s^23p^63d^24s^2$ (Ti), and $2s^22p^4$ (O).[23,24] An effective Hubbard term $U_{eff} = U–J$ using Dudarev's approach with $U_{eff}$ = 3.27 eV was included to treat the Ti 3*d* orbital,[25] and a 520 eV cutoff energy of the plane-wave basis set was used for all calculations. For structure optimizations, atomic positions were considered relaxed for energy differences up to $1 \times 10^{-6}$ eV and all forces were smaller than 1 meV Å$^{-1}$. During the relaxation of bulk PbTiO$_3$ unit cell (u.c.) with electronic electrostatic doping concentration 0.12 *e*/u.c., 11 × 11 × 11 gamma-centered *k*-point meshes were used, while the doping electrons were introduced to the system with a uniform and positive background to achieve charge neutrality. Denser k-point meshes were used for density of states calculations.

The atomic structures and isosurfaces of the charge density were visualized using the VESTA package.[26]

## 3. RESULTS AND DISCUSSION

PbNb$_{0.12}$Ti$_{0.88}$O$_3$ (PNTO) thin films were grown on (001)-oriented SrTiO$_3$ (STO) and Nb:SrTiO$_3$ (SNTO) substrates using a Laser-MBE system. The results of X-ray diffraction (XRD) and X-ray diffractometry reciprocal space mapping (RSM) indicate that the PNTO films have high crystalline quality and possess a single phase (Supporting Information, Figure S1). X-ray absorption spectra (XAS) were collected in the PNTO films under the total electron yield (TEY) mode, suggesting that the as-grown films were free of oxygen vacancies (Supporting Information, Figure S2).[27,28]

Figure 1a shows the representative high-angle annular dark-field scanning transmission electron microscopy (HAADF-STEM) image of a PNTO film grown on a STO substrate. The STEM images demonstrate that the PNTO film is epitaxial growth, free of defects, having a sharp film-substrate interface (Supporting Information, Figure S3), and that no obvious oxygen vacancies were observed in the PNTO films (Supporting Information, Figure S4).[29] An atomic-resolved STEM image is shown in Figure 1b. The intensity of STEM image is proportional to $Z^2$, where the $Z$ is the atomic number of elements.[5] Therefore, the heavier atomic columns (Pb) manifest the brighter contrast, while the lighter ones (Nb/Ti) are darker. The conspicuous displacements of Pb atoms were observed in the atomic-resolved STEM image. As indicated by the yellow arrow in Figure 1b, the Pb atoms shift upwards from the center of Nb/Ti cubic, indicating the intrinsic polarization is along the out-of-plane direction. We statistically analyzed the atomic displacements of the Pb atoms with respect to the Nb/Ti atoms quantitatively, yielding an average value of 0.250±0.068 Å (Figure 1c).[8] The STEM results indicate the microscopic origin of polar structure in the PNTO films.

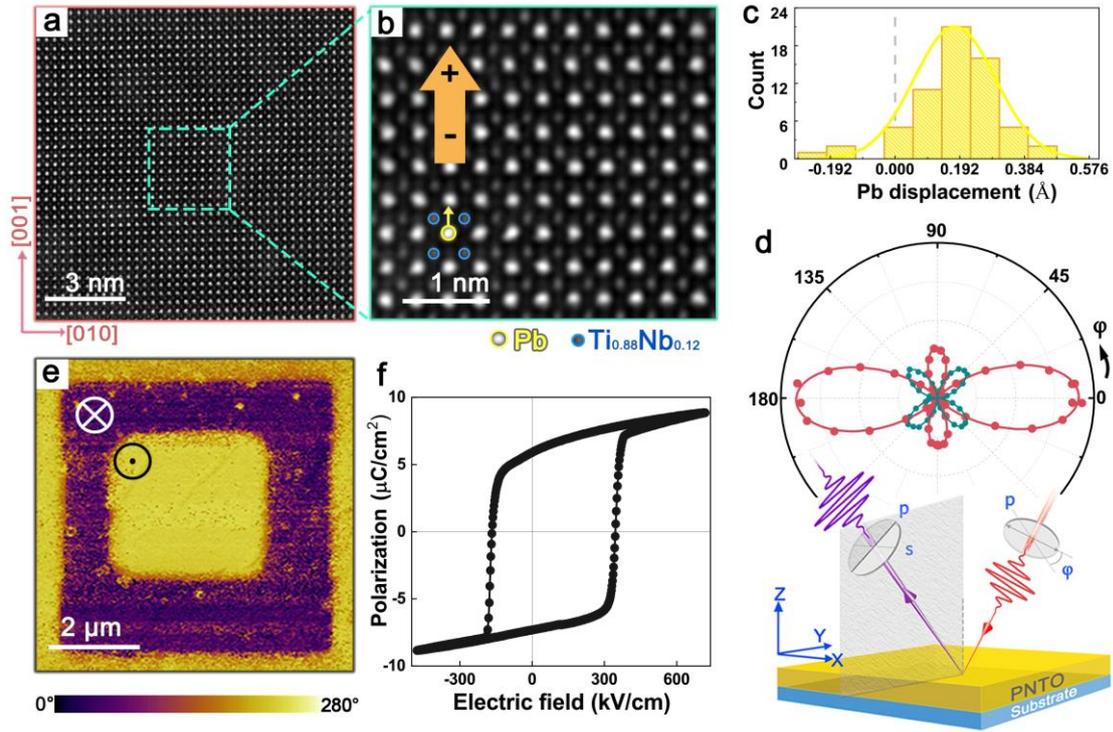

**Figure 1.** Ferroelectric nature of the heavily electron-doped PNTO single films. (a) HAADF-STEM image of a PNTO film. The zone axis is along the [100] direction of the STO substrate. (b) Atomic-resolution STEM image of a selected area marked by the dashed box in (a). The atomic Pb displacement is illustrated relatively to the positions of the $Nb_{0.12}Ti_{0.88}$ atoms. The yellow arrow indicates the Pb atom displacement and the orange arrow shows the polarization direction. (c) Statistical distribution of Pb displacement along [001] direction shown in (b). (d) Polar plot of the polarization-dependent SHG signals for a PNTO film. Two SHG components, *p*- and *s*-out, are plotted with analyzers parallel or perpendicular to the incidence plane, respectively. The solid circles are experimental data and the solid lines are theoretical fits. The inset shows schematic diagram of the reflective SHG measurements on a PNTO sample. (e) PFM phase image obtained from a PNTO thin film. Double squares illustrate the areas switched by a conductive tip with a polarization pointing upwards for the smaller square (3 × 3 $\mu m^2$) and downwards for the bigger square (6 × 6 $\mu m^2$). (f) *P-E* hysteresis loop of a PNTO film, indicating a maximum remnant polarization of ~6.5 $\mu C/cm^2$. All measurements are conducted at room temperature.

The polar structure of PNTO films was further confirmed by reflected SHG measurements (Figure 1d). The SHG anisotropic patterns were obtained in *p*-out and *s*-out configurations.[8,9,18] Theoretical fits of the SHG results confirm and indicate that the point group of the PNTO film is *4mm*,[18-20] identical with the structure obtained from our first principle calculations (Supporting Information, Figure S5). The structural characterizations show that the PNTO films possess non-trivial polar structure. We do not exclude the contribution from the surface and the interface to the SH signals of the films, but that should be comparatively small concluded from our measurements for much thinner films.

The switchable ferroelectric polarization in the heavily electron-doped PNTO films was further confirmed by PFM and macroscopic electrical measurements. Figure 1e shows the phase contrast of a PFM image on a PNTO thin film. The as-grown PNTO film stabilizes in a single-domain self-poled state with upward ferroelectric polarization. This observation agrees with the STEM results (Figure 1b). Firstly, we applied a positive bias through a conductive tip in a large area (6×6 $\mu m^2$), and then an inner square (3×3 $\mu m^2$) was switched backwards using a negative voltage. The color contrast in the PFM image indicates the ferroelectric domains with opposite polarization. The sharp contrast between the up and down ferroelectric domains demonstrates the good microscopic ferroelectric character. Figure 1f shows the *P-E* hysteresis loop measured in a PNTO thin film. The distinguishable square-like curve indicates the intrinsic ferroelectric characteristic of the film, yielding a remnant polarization as high as ~6.5 $\mu C/cm^2$. The off-centered *P-E* loop to the positive side of the Electric field axis indicates that the polarization in the as-grown PNTO thin films points to the sample surface,[30] which is concordant with the STEM results in Figure 1b. These results indicate that the heavily doped PNTO films not only exhibit an intrinsic polar structure,[5] but also have the measurable microscopic and macroscopic ferroelectric characteristics, as the electric polarization is fully reversible by an electric field with opposite signs.

The electrical transport properties of the PNTO films were characterized using both transverse and longitudinal measurement geometries, as shown in the insets of Figures 2a and 2b, respectively. In the transverse geometry, the resistivity keeps decreasing as the temperature increasing from 10 K to room-temperature, which exhibited a semiconducting (or insulating) behavior along the in-plane direction of PNTO films. However, in the longitudinal geometry, after firstly decreasing, the resistivity increases with the temperature, which demonstrated the transition of semiconducting (or insulating) to metallic behavior in the out-of-plane direction of PNTO films. The strong anisotropy in the electrical conductivity is explained qualitatively by our first principle calculation (Supporting Information, Figure S5), and more universal predictions about that in polar metal can be found in our previous work.[13,31] As seen in Figure 2b, the PNTO films undergo a metal-to-insulator transition at ~190 K. This temperature corresponds to the phase transition of $PbTiO_3$ at low temperatures.[16]

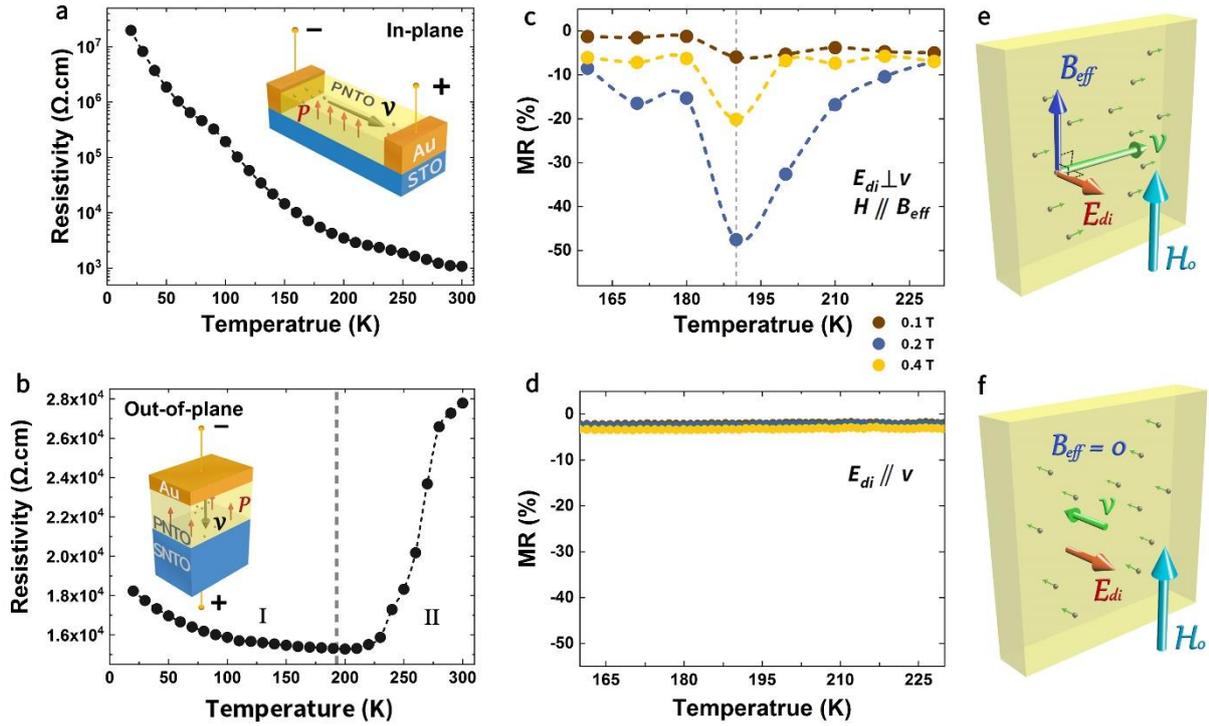

**Figure 2.** Asymmetric resistivity and MR in the PNTO films. The solid circles in (a-d) are experimental data. (a) and (b) Resistivity of the PNTO films measured in the transvers and longitudinal geometry, respectively. Insets illustrate the schematics of measurement geometries. (c) and (d) Temperature dependent MR measured in the transvers and longitudinal geometry, respectively. The external magnetic field $\vec{H}$ is along the in-plane direction, which is perpendicular to the electric polarization of the PNTO films. A small current of 1 μA is applied for all the electrical measurements. Negative MR is observed for the transvers geometry; while the MR is nearly zero for the longitudinal geometry. (e) and (f) Schematics of the effective magnetic fields ($\vec{B}_{eff}$) produced in transvers and longitudinal geometry, respectively. The grey balls stand for conductive electrons.

The intriguing magnetoresistance (MR) was observed as shown in Figure 2c, calculated by MR

= {[$R(H)$-$R(0)$]/$R(0)$}×100%, with $R(H)$ and $R(0)$ being the measured resistivities of the PNTO films under and without external magnetic fields, respectively. A clear negative MR with the maximum up to 50% occurred at ~190 K under an external magnetic field of 2 kOe in the transverse geometry. This MR in the PNTO films is significantly larger than the ordinary MR (less than 2%) reported in the normal metals or semiconductors induced by the conventional Hall Effect,[32,33] but usually be found in ferromagnetic materials.[34-38] To verify whether the PNTO films possess any ferromagnetism or not, we conducted the magnetization measurement about the as-grown PNTO films without external electric field, and the results (Supporting Information, Figure S6) confirmed that these films are diamagnetic, which excluded the intrinsic ferromagnetic origin for the observed MR.

In sharp contrast to the conventional metals and ferroelectrics, the present PNTO films possess both electrical polarization and mobile electrons. In a primitive cell of PNTO, the center of negative charges separates from that of positive ones, forming an electric dipole. This dipole generates a localized electric field $\vec{E}_{di}$. When an electron moves perpendicular to the dipole induced electric field $\vec{E}_{di}$ driven by an applied electric field, according to the interpretation of electric and magnetic fields transformation[39], this relative movement between the electron and the dipole induced electric field can generate an effective magnetic field, written as:

$$\vec{B}_{eff} = \frac{\vec{E}_{di} \times \vec{v}}{c^2}, \qquad (3)$$

where $\vec{v}$ is the relative velocity of $\vec{E}_{di}$ with respect to the electron, and $c$ is the speed of light in vacuum. The direction of the effective magnetic field $\vec{B}_{eff}$ is perpendicular to the plane determined by $\vec{v}$ and $\vec{E}_{di}$, schematically shown as Figure 2e. The $\vec{B}_{eff}$ is the effective magnetic field that one electron feels while moving in the PNTO films,[39] which may couple with the spin of the moving electron. In the case that this effective magnetic field couples with the spin of the moving electrons, a magnetic ordering in the film can be generated. Since this kind of magnetic ordering is only accompanied by the directional movement of electrons—it will vanish as long as the moving stops, we note this temporary ordering as *transient magnetic ordering* and corresponding state of PNTO films as the *transient ferromagnetic state*. This transient magnetic ordering exists as long as the electrons vectored move or the external electric field is applied perpendicular to the $\vec{E}_{di}$. The experimental results in Figure 2c indicate such a transient magnetic ordering in the PNTO films, which converted the film from an initial diamagnetic state to a "*transient ferromagnetic*" state. With an externally applied magnetic field $\vec{H}$, parallel to $\vec{B}_{eff}$, the "*transient ferromagnetic*" domains merged with one another, largely reducing the electron spin scattering, leading to the observed negative MR. This negative MR has been widely found in some ferromagnetic materials, like manganites,[34,38,40] cobaltites,[41,42] and etc.

From Figure 2a, we can also see that the resistivity of PNTO films decreased sharply with the temperature increasing, indicating a great increase of the electron mobility ($\vec{v}$). Meanwhile, the polarization of PNTO films decreased largely with the temperature rising, which can be concluded from our observation in the SHG measurement,[28] resulted in a large decrease of induced electric field $\vec{E}_{di}$ in films. This competition between the increase of $\vec{v}$ and the decrease of $\vec{E}_{di}$ with increasing temperature might result in the effective magnetic field $\vec{B}_{eff}$ reaching a maximum, and further influence the MR reaching a peak value around 190 K. It should be noticed that this critical temperature is also the phase transition temperature for PTO in low temperature region,[16] thus the phase transition might also be a reason for the appearance of the MR maximum. Electric field-dependent MR of PNTO films are also shown for various temperatures, 160, 180, 200, and 260 K (Supporting Information, Figure S7).

To verify the mechanism that the negative MR was caused by $\vec{B}_{eff}$, the contrastive experiment was carried out. At this time, the MR was measured in the longitudinal geometry with the external magnetic field applied in the same in-plane direction as above (Figures 2d and 2f). As we can see, the MR measured is negligibly small (~2%) and is independent with temperature (Figure 2d), which proved the above mechanism directly. As the electrons move alone the direction of the polarization ($\vec{v} \parallel \vec{E}_{di}$), there is no effective magnetic field produced, meaning $\vec{B}_{eff} = 0$. The none zero value of MR, or none zero $\vec{B}_{eff}$, in our observation could come from the small in-plane component of ferroelectric polarization in PNTO films.

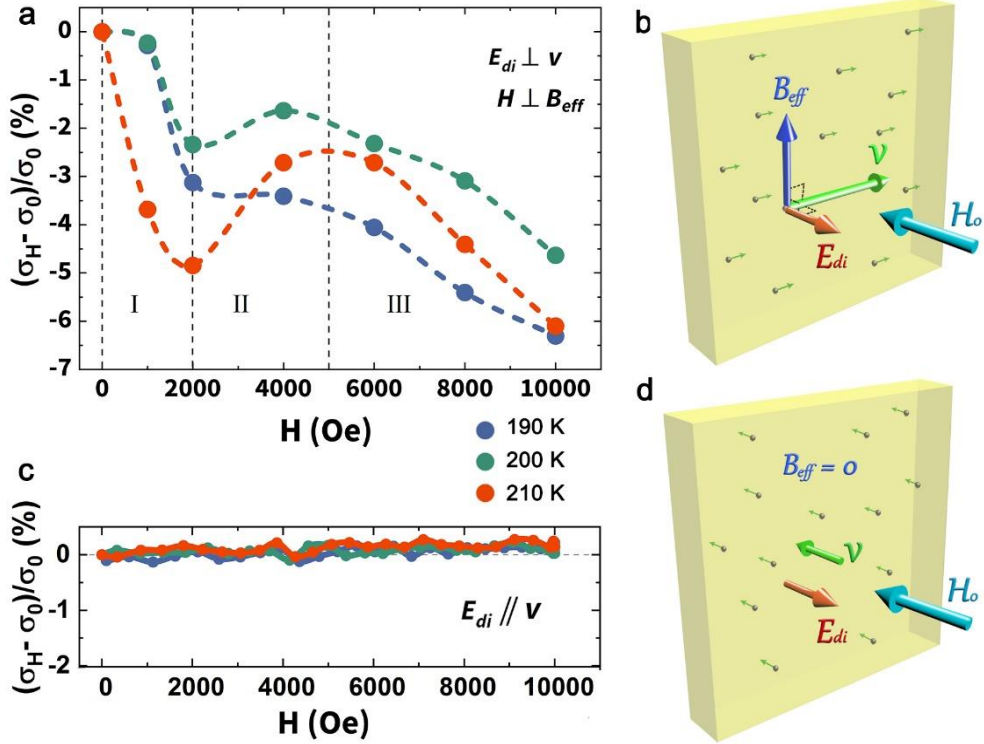

**Figure 3.** Positive MR in the PNTO films. (a) $\vec{H}$-dependent in-plane magnetoconductance changing when $\vec{H}$ is perpendicular to $\vec{B}_{eff}$. The dashed lines are for eye guide in (a). (b) Schematic diagram of the measurement for (a). (c) $\vec{H}$-dependent out-of-plane magnetoconductance changing when $\vec{H}$ is perpendicular to $\vec{B}_{eff}$. (d) Schematic diagram of the measurement for (c). The grey balls in (b) and (d) stand for conductive electrons.

To further prove the strong coupling of effective magnetic field and spins of electrons, we changed the direction of external magnetic field, making it parallel to the surface normal ($\vec{H} \parallel \vec{E}_{di}$), the MR of the PNTO films turned to be positive (Figures 3a and 3b), correspondingly. With rising $\vec{H}$, the MR increased (magnetoconductivity decreased) in the low fields ($\vec{H}$ < 2 kOe, region I), and reaches its maximum (minimum) at ~2 kOe; then the MR decreased (magnetoconductivity increased) in the middle field (region II). In the high fields ($\vec{H}$ > 5 kOe, region III), the MR increased (magnetoconductivity decreased) again with rising $\vec{H}$.

As far as we know, there is no temperature-dependent MR changing like this in any films with the thickness of hundreds of nanometers. However, such situation can be widely observed in the 2D systems and ultrathin perovskite superlattice,[43-48] which were explained by weak localization (WL) corrected by quantum mechanics.[49,50] According to the WL theory, in low field region, electrons transporting in weak-disorder systems experience many elastic and inelastic scattering events. When the inelastic scattering length is larger than the elastic scattering length, the coherent back scattering can be enhanced by the constructive interference of electrons propagating in time-reversal closed trajectories in the absence of spin-orbit scattering and magnetic scattering, leading to a negative quantum correction to the classical conductivity.[51] With the field increasing, the applied magnetic field will induce an additional phase shift that will suppresses WL and produces a positive magnetoconductivity.[51] With the field continually increasing, ordinary MR caused by the Lorenz force begin to play the dominate role in the system.

The similarities of the PNTO films and the above systems are that their symmetry are all broken out-of-plane, the external magnetic fields $\vec{H}$ are both applied out-of-plane, and the electrons moving/current are in-plane. The differences are the much higher temperature and much lower $\vec{H}$ for PNTO films (~190 K, 1 T) to other systems (~10 K, ~8 T). This unconventional MR should, we suppose, due to the coexistence of electrical polarization and mobile electrons in a single material.

The contrastive experiment was also carried out, in which the MR was measured in the longitudinal geometry as the external magnetic field applied perpendicular to the film surface (Figures 3c and 3d). Negligible MR was observed for all values of the applied field, further verified that the effective magnetic field can cause a transient magnetic ordering coupling with the spins of electrons as long as electrons move perpendicular to the polarization.

This transient magnetic ordering formed in the ferroelectrics with electron moving makes this material in a *dynamic multiferroic* state. This observation would further stimulate the investigation of the multiferroic metal and the quantum perturbation of polar metals at high temperatures.

## 4. CONCLUSION

In summary, we report the observation of the extraordinary magnetoresistance in the metallic ferroelectric PNTO films for the first time. The structural characterizations obtained from the combined results of STEM, XRD, and SHG confirm the polar structure in the stoichiometric PNTO films. PFM and electrical measurements show that the electrical polarization in the heavily electron-doped PNTO films can be effectively switched by applied electrical field. Under the external magnetic fields applied parallel to the sample surface, the non-ferromagnetic PNTO films exhibit a negative magnetoresistance up to 50% at ~190 K. The positive magnetoresistances in the PNTO films were observed when the external magnetic field is applied parallel to the surface normal, further verifying the coupling of effective magnetic field with spins of electrons. We attribute such intriguing magnetic properties to the *"ferromagnetic-like"* state induced by the effective magnetic field, with electrons moving in the direction perpendicular to the intrinsic electrical polarization, coupled with the spins of electrons. Meanwhile, we suggest a new concept of *dynamic multiferroicity*, that is, a transient magnetic ordering formed in metallic ferroelectrics, as long as electrons move perpendicular to the polarization. We hope these findings can pave a revolutionary avenue towards the multifunctional devices.

**AUTHOR INFORMATION**

Corresponding Author

*E-mail: kjjin@iphy.ac.cn

**ORCID**

Kuijuan Jin: 0000-0002-0047-4375

Can Wang: 0000-0002-4404-7957

**Author Contributions**

‖J.S.W. and H.B.Y. contributed equally to this work. K.J.J. conceived the project. J.S.W. performed the ferroelectric and SHG measurements. H.B.Y. performed the electrical and magnetic measurements. Q.H.Z. and G.L. did the STEM experiment. J.S.W., H.B.Y., and V.P. prepared the PNTO films. C.M. carried out the theoretical calculations. J.L.Z and J.O.W performed the XAS measurement. K.J.J. supervised the experimental and theoretical studies. E.J.G., K. J. J., J.S.W., C.M, H.B.Y., and Q.H.Z. wrote the manuscript. All authors discussed the results and commented on the manuscript.

**Notes**

The authors declare no conflict of interest.

**Supporting Information**

Structural characterization of the fabricated films, X-ray absorption spectra, local structural characterization, oxygen vacancy analysis, theoretical calculation of asymmetrical electron distribution in $PbTiO_3$ with 0.12 $e$/u.c. doping, M-H property, and magnetoresistance as a function of applied voltage.

**ACKNOWLEDGMENTS**

This work was supported by the National Key Basic Research Program of China (Grants No. 2017YFA0303604), the National Natural Science Foundation of China (Grants Nos. 11721404, 51761145104, and 11674385), the Key Research Program of Frontier Sciences of the Chinese Academy of Sciences (Grant No. QYZDJ-SSW-SLH020), and the Strategic Priority Research Program (B) of the Chinese Academy of Sciences (Grant No. XDB07030200).

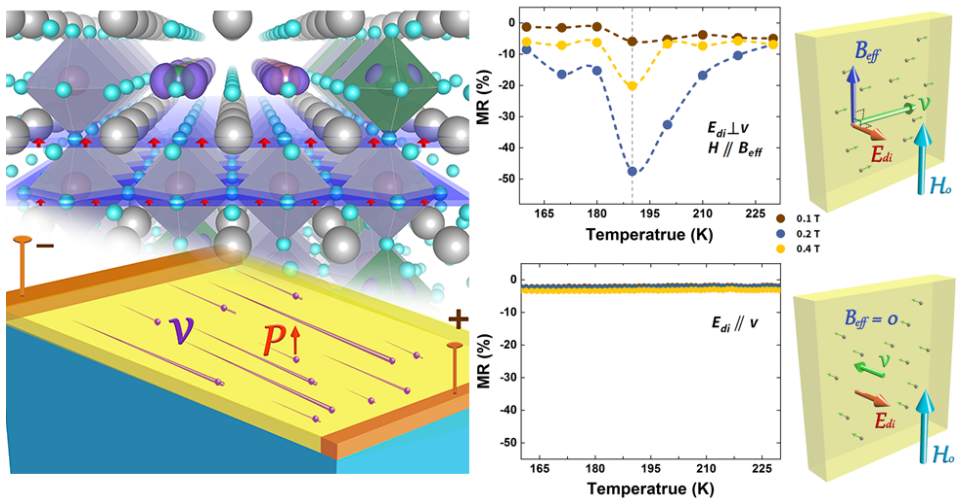

TOC Graphic